\documentclass[aps,prr,twocolumn,floatfix,amssymb]{revtex4-2}

\usepackage{graphicx,dcolumn,bm,xstring,braket,amsmath,comment,nicefrac,placeins,multirow}
\usepackage[unicode=true,pdfusetitle,bookmarks=true,bookmarksnumbered=true,bookmarksopen=false, breaklinks=false,pdfborder={0 0 0},pdfborderstyle={},backref=false,colorlinks=false, pdftitle={Atomic interferometer based on optical tweezers}]{hyperref}

\newenvironment{DIFnomarkup}{}{}

\begin{document}

\title{Atomic interferometer based on optical tweezers}

\author{Jonathan Nemirovsky} 
\author{Rafi Weill}
\author{Ilan Meltzer}
\author{Yoav Sagi}
\email[Electronic address: ]{yoavsagi@technion.ac.il}
\affiliation{Physics Department and Solid State Institute, Technion - Israel Institute of Technology, Haifa 32000, Israel}

\date{\today}

\begin{abstract}

Atomic interferometers measure forces and acceleration with exceptional precision. The conventional approach to atomic interferometry is to launch an atomic cloud into a ballistic trajectory and perform the wave-packet splitting in momentum space by Raman transitions. This places severe constraints on the possible atomic trajectory, positioning accuracy and probing duration. Here, we propose and analyze a novel atomic interferometer that uses micro-optical traps (optical tweezers) to manipulate and control the motion of atoms. The new interferometer allows long probing time, sub micrometer positioning accuracy, and utmost flexibility in shaping of the atomic trajectory. The cornerstone of the tweezer interferometer are the coherent atomic splitting and combining schemes. We present two adiabatic schemes with two or three tweezers that are robust to experimental imperfections and work simultaneously with many vibrational states. The latter property allows for multi-atom interferometry in a single run. We also highlight the advantage of using fermionic atoms to obtain single-atom occupation of vibrational states and to eliminate mean-field shifts. We examine the impact of tweezer intensity noise and demonstrate that, when constrained by shot noise, the interferometer can achieve a relative accuracy better than $10^{-11}$ in measuring Earth's gravitational acceleration. The sub-micrometer resolution and extended measurement duration offer promising opportunities for exploring fundamental physical laws in new regimes. We discuss two applications well-suited for the unique capabilities of the tweezer interferometer: the measurement of gravitational forces and the study of Casimir-Polder forces between atoms and surfaces. Crucially, our proposed tweezer interferometer is within the reach of current technological capabilities.
\end{abstract}

\maketitle


\section{Introduction}

Interferometers have a long history of driving scientific revolutions, from the Michelson-Morley experiment \cite{michelson1887relative}, to the recent observation of gravitational waves with the Laser Interferometer Gravitational-Wave Observatories (LIGO, Virgo and KARGA) - the most advanced interferometer ever built \cite{Abbott2016}. Soon after the discovery of the wave-particle duality, in the early years of the 20th century, it was realized that the interference of massive particles could be harnessed for the purpose of highly precise measurements \cite{Estermann1930,Berman1997}. Over the years, matter-wave interference has been demonstrated using a wide range of masses, including electrons, atoms, and complex molecules \cite{Cronin2009,Schaff2014}. The development of laser cooling techniques has made cold atoms a popular choice for interferometry due to their large de Broglie wavelengths and slow velocities, which allow for long coherence and integration times. 

Atomic interferometers (AIF) come in many forms, but they all rely on the same fundamental principle; the atomic wave packet is initially prepared in a specific state and then coherently divided into two parts that follow distinct paths \cite{Berman1997}. The quantum wave function in each arm may acquire a different phase. The two arms are then coherently combined, and the process of recombination maps the relative phase shift between them to populations in two output states, which may be external (e.g., spatial modes, momentum states) or internal (e.g., spin projections, atomic energy levels). The most significant distinction between atomic and photonic interferometers is the non-zero mass of the former. This means that in an atomic interferometer, atoms can be brought to a complete halt.

\begin{figure*}
 \centering
	\includegraphics[width=\textwidth]{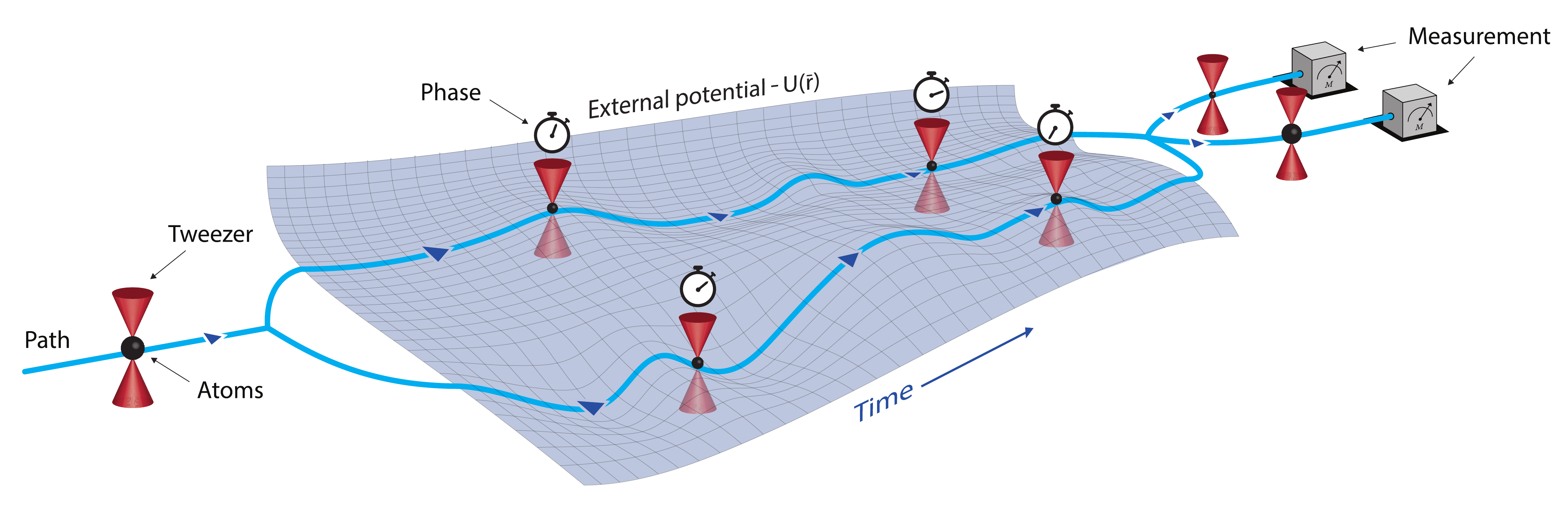}
	\caption{\textbf{Tweezer-based atomic interferometer.} Fermionic atoms are held by an optical tweezer, occupying the lowest vibrational eigenstates. The tweezer splits and recombines coherently the atomic wave packet, with each part of the wave packet accumulating a different phase as it travels in a different path through the external potential. The combiner translates the relative phase shift between the arms into a difference in population (marked by the size of the black sphere). The atoms can be trapped for many tenths of seconds, and the path can be shaped with sub-micrometer resolution, giving the interferometer unparalleled precision and flexibility.}	\label{fig:graphical_abstract}
\end{figure*} 

To coherently split wave-functions of atoms, AIF have initially employed diffraction from periodic fabricated structures \cite{Keith1988,Keith1991} and optical lattices \cite{Moskowitz1983}. These techniques utilize the exchange of lattice momentum to create sidebands in the atomic wave-function's momentum distribution, which leads to a coherent splitting of the atomic path in real space. They are relatively simple to implement and robust, but they have low efficiency and limited control over the atomic trajectory. A different approach was to utilize the coherent absorption of a single photon \cite{Borde1989} or two photons with different wave-vectors \cite{Chebotayev1985} to generate coherent splitting in momentum space. One well-known example is the Kasevich and Chu (KC) interferometer \cite{Kasevich1991}, which uses stimulated Raman transitions to drive coherent Rabi oscillations between two momentum states, which then undergo different kinematic trajectories. Stopping the oscillation after quarter of a period ($\nicefrac{\pi}{2}$ pulse) generates a balanced superposition of these states. The KC interferometer is based on a sequence of $\nicefrac{\pi}{2}$-$\pi$-$\nicefrac{\pi}{2}$ pulses to achieve the splitting, mirroring and recombining of the atoms, respectively. 

KC interferometers were instrumental for many precision measurements in the last two decades, including determination of the gravitational constant \cite{Fixler2007,Rosi2014}, measurement of the fine structure constant \cite{Parker2018}, testing the equivalence principle \cite{Asenbaum2020}, and constraining dark energy models \cite{Hamilton2015}, to name a few. However, they suffer from several shortcomings, including limited spatial resolution, and atomic motion which is geodesic only (i.e., free-fall) and cannot be freely shaped. In particular, it is not possible to position the atoms at rest at arbitrary locations. Moreover, to have a long probing duration, the experimental apparatus tend to be very large, and, even then, the interaction time is limited to few seconds. AIF with three-dimensional confinement of the atoms would be a complete change of paradigm \cite{Raithel2022}. A step in this direction was recently reported by the group of H. Muller, who developed a variant of a KC interferometer that combines trapping the atoms at the apex of the geodesic motion, reaching a holding time of 20-60 seconds \cite{Xu2019,Panda2022}. However, the atomic motion was still ballistic and the maximum separation between the wave packets was tens of micrometers. Mater wave interference with 3D-confined condensates of bosons and fermionic pairs was also demonstrated, but only as a tool to study the coherence of the condensate wave function \cite{Shin2004,Valtolina2015}. Specifically, substantial stochastic phase shifts due to inter-particle interactions in these gases make them unsuitable for precision metrology. Interferometry with a single trapped atom was demonstrated in a spin-dependent lattice \cite{Steffen2012}. However, the maximum separation was around 10 micrometers and the holding time was limited to around 1 ms due to spontaneous scattering from the lattice.

Here, we propose a new scheme for atomic interferometry that employs mobile micro-optical traps, known as ``optical tweezers'', to trap and manipulate individual atoms. Recent years have seen significant progress in this field \cite{Kaufman2021,Graham2022,Bluvstein2022,Spar2022,Yan2022,Menchon-Enrich2014}. Neutral atoms in optical tweezers have been used in quantum computing as qubits \cite{Graham2022,Bluvstein2022}, in quantum simulation of many-body phenomena \cite{Browaeys2020,Spar2022}, and for precision time measurements \cite{Norcia2019}. We propose to use the tweezers to coherently split and recombine the atoms, and in between to hold the atomic wave packets for tens of seconds with sub-micrometer positioning accuracy and complete freedom to shape the atomic trajectory. A key aspect of our proposal is the implementation of atomic splitters and combiners that do not change the internal state of the atom, are robust to experimental imperfections, and work with many vibrational states of the tweezer. Furthermore, we propose to use fermionic atoms and leverage their Fermi-Dirac statistics to have between few tens to a hundred atoms in a single run while avoiding systematic interaction energy shifts. This unique combination will allow high-precision measurement of potentials with sub-micrometer resolution.

The structure of this paper is as follows. In Sec. \ref{sec:tweezer-based atomic interferometer} we describe the tweezer AIF in detail. In Sec. \ref{sec:Coherent splitting and recombining}, we discuss the atomic splitting and recombining schemes. We first explain the general considerations, and then present two schemes based on two and three tweezers. In Sec.~\ref{sec:Sensitivity_estimation} we present numerical simulations done with realistic noise parameters in order to estimate the sensitivity and precision of the proposed interferometer. In Sec. \ref{sec:Applications} we discuss two physical measurements where the new AIF can be particularly beneficial:  measurement of the gravitational constant and measurement of Casimir-Polder forces. We summarize and give an outlook in Sec. \ref{sec:Summary}.

\section{tweezer-based atomic interferometer}\label{sec:tweezer-based atomic interferometer}

A schematic sketch of the new interferometer is shown in Fig. \ref{fig:graphical_abstract}. An atomic wave-packet is prepared and held in a single optical tweezer. Then, the wave-packet is split coherently into two tweezers, each moving in a different path, and afterwards it is recombined. By interfering the two wave-packets, one can detect the relative phase shift between the arms, which arises due to differences in the external potential and dynamics along the paths. Typically, experiments with ultracold atoms are conducted in an ultra-high vacuum chamber, allowing to hold the atoms for a very long duration (i.e., tens to few hundreds of seconds). Furthermore, optical tweezers can have a Gaussian waist of around $1\mu$m, in which the atomic wave-function is typically localized to around 200 nm. Combining this with a positioning precision of few hundreds of nanometers yields a spatial resolution better than a micrometer. Two central aspects of the tweezer interferometer are the adiabatic atomic splitter-combiner and the use of identical fermions. The fermionic statistics guarantees that each energy eigenstate of the tweezer ('vibrational' states) is occupied by at most a single atom and that the atoms do not interact.

To achieve these conditions, the tweezer can be loaded from a moderately degenerate Fermi gas ($T/T_F\approx 1$, where $T_F$ is the Fermi temperature), harnessing the ``dimple effect'' to enhance the phase space density \cite{StamperKurn1998,Serwane2011}. Thus, the occupation probability in all low-lying eigenstates can be very close to unity. After loading the tweezer, the atoms occupying the highest eigenstates are eliminated by gradually reducing the trap depth, until the desired number of atoms is reached \cite{Serwane2011}. The Pauli exclusion principle ensures that at each vibrational state there is at most one atom at a specific spin state. Note that to operate the interferometer with many atoms, better initial conditions of the initial degenerate Fermi gas will be required. As for the question of the atomic species, there is an advantage to work with atoms with a higher mass, since the interferometer will be more sensitive to acceleration and gravitational potential. In the alkali group, $^{40}$K is preferable, and in the lanthanide group, $^{171}$Yb is a promising candidate \cite{Ma2022,Jenkins2022}.

It is desirable to operate the tweezer with as many atoms as possible to reduce the number of repetitions needed to achieve a certain level of uncertainty. Additionally, having many atoms in a single run allows one to measure transient phenomena which cannot be averaged. Our approach to achieve this is to utilize many vibrational states in a single tweezer. In Sec. \ref{sec:Coherent splitting and recombining} we present the splitting and recombining schemes and show that, with the same set of parameters, they work successfully for many vibrational states. This property allows for multiple atoms to be used in the interferometer at the same time \cite{Andersson2002}. Moreover, the combiner maps differential phase shifts between the interferometer arms to population differences in the output ports in a manner that does not depend on the vibrational state. In this way, a single experimental run using $N$ atoms in a single tweezer is equivalent to using $N$ tweezers with a single atom each. We estimate that realistically, 100 atoms can be used in each run, yielding a 10-fold improvement in the signal-to-noise ratio compared to a single-atom interferometer. Additionally, the splitting does not change the atoms' internal state, making the superposition robust to spin-dependent noise. Importantly, fermionic anti-symmetry precludes interaction between the identical atoms, which avoids systematic shifts. 

\section{Coherent splitting and recombining}\label{sec:Coherent splitting and recombining}
A crucial part of the new interferometer is the coherent splitting and merging of the atomic wave-packet. We present here two approaches to achieve this, using two and three tweezers. A tweezer interferometer based on the two-tweezer scheme is somewhat similar to the optical Mach–Zehnder interferometer. Our three-tweezer splitter-combiner scheme, on the other hand, has no optical analog, to the best of our knowledge. Both schemes are robust to experimental imperfections, but the one based on three tweezers also allows for the detection of errors, a unique capability that does not exist in any other photonic or atomic beam splitter.

Adiabatic splitting and recombining scheme is based on the idea that the initial state, where atoms are localized in one of the tweezers, and the final state, where the atoms are in a balanced coherent superposition of being in two tweezers, are continuously connected through an adiabatic change of some external parameters. Adiabatic driving is widely used to manipulate internal states of atoms and molecules. The version applied to a 3-state lambda configuration, known as stimulated Raman adiabatic passage (STIRAP) \cite{Gaubatz1990}, has found wide-ranging applications in many fields of science \cite{Vitanov2017}. The concept of STIRAP has also been extended to tunneling between three spatially separated potential wells \cite{MenchonEnrich2016}. Spatial adiabatic passage (SAP) has been recently demonstrated experimentally with fermionic $^{40}$K atoms in optical tweezers \cite{Florshaim2023}. The mapping of STIRAP to a two level system was explored in Ref.~\cite{Vitanov2006}. The notion that atomic interferometry can be based on a SAP protocol was introduced in Ref.~\cite{Menchon-Enrich2014}. The schemes for splitting and combining that we present below draw inspiration from these aforementioned works.

There are several characteristics that we aim to achieve with the atomic splitter; \emph{Reversibility} -- The process should involve at least two input and output ports, and it should be reversible, meaning that application of the scheme followed by its time-reversed version brings the atoms back to the original port. \emph{Ease of detection} --  Occupation in the different output ports should be easy to detect. Specifically, it is much easier to detect the population in two spatially separated tweezers than to distinguish between the population in two different vibrational states of the same tweezer. \emph{Robustness against experimental imperfections} -- The process should withstand small variations in parameters, such as duration, trap intensity, and position. Since we want to use multiple atoms occupying different vibrational states in parallel, it should ideally also be insensitive to the vibrational state.

To illustrate the importance of our adiabatic driving schemes, let us consider first a simple non-adiabatic $\nicefrac{\pi}{2}$ splitter. The initial state is one tweezer occupied by a single atom in some vibrational eigenstate and a second empty tweezer at a close proximity such that there is tunneling. As time progresses, the atom undergoes tunneling oscillations back and forth between the two tweezers \cite{Kaufman2014}. If the coupling between the tweezers is terminated exactly in the midst of such an oscillation (e.g., by moving the tweezers apart), the atomic wave packet will be coherently divided between the two tweezers. This splitting scheme is analogous to a $\nicefrac{\pi}{2}$ pulse in Rabi oscillations. However, it has several disadvantages. First, the splitting is first-order sensitive to changes in the process duration. Second, since the tunneling rate is strongly dependent on the distance, any positioning noise is translated into fluctuation in the splitting process. The strong distance dependence also means that the tunneling rate depends on the vibrational eigenstate, hence this scheme can work only for a single state. The two schemes we present below follow adiabatic driving and are therefore robust to experimental imperfections and vibrational state occupation.

\subsection{Two-tweezers atomic splitter-combiner}\label{sec:Two-tweezers atomic splitter}

 \begin{figure*}
	\centering
\includegraphics[width=1\textwidth]{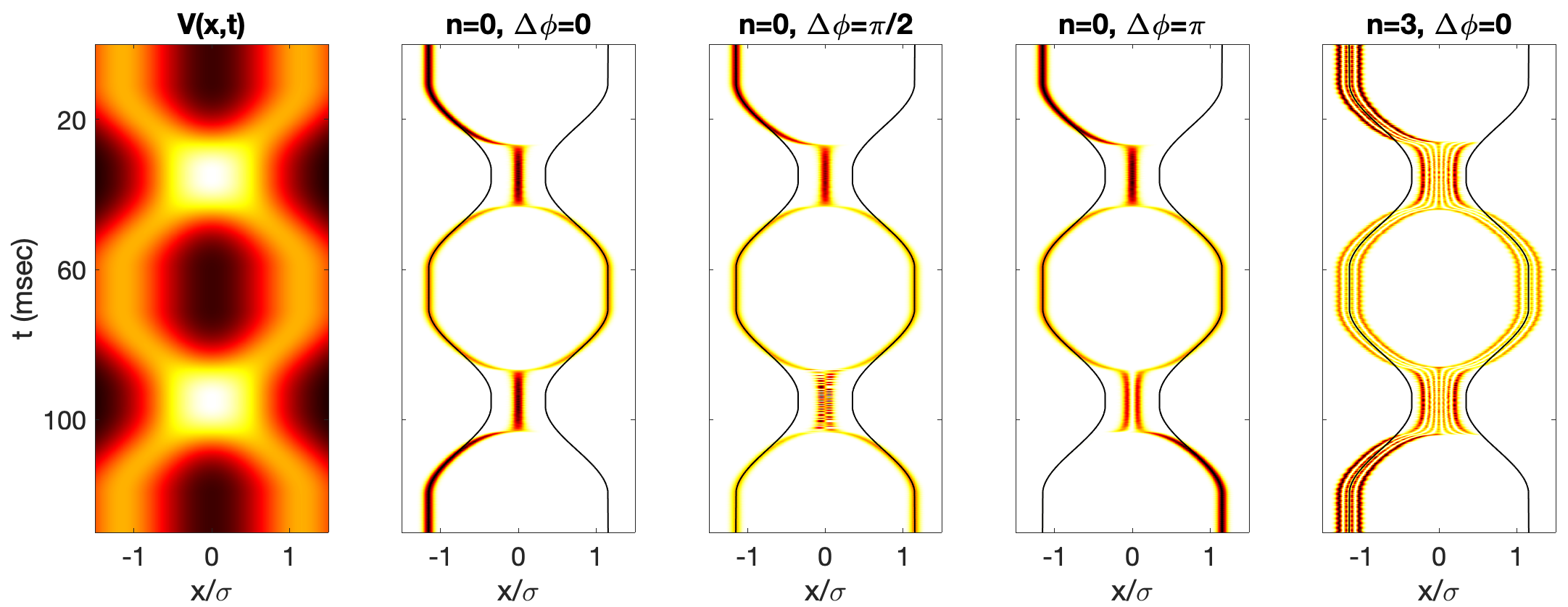}
\caption{\textbf{Numerical simulation of the two-tweezer interferometer with a $^{40}$K atom.} On the left, the optical potential is plotted as a function of position (horizontal) and time (vertical). The tweezers have a waist of $\sigma=1.3\mu$m and a final depth of $116\mu$K. The right tweezer starts with a detuning of $\Delta=-2.3\mu$K. The other four panels show the probability distribution $|\psi(x)|^2$ (represented by the brightness of the color) with different initial vibrational state $n$ and different relative phases $\Delta \phi$ between the arms. The black lines mark the tweezers paths. One can see that the splitting works regardless of the vibrational state, and that the relative phase is translated into population difference in the output ports.}	\label{fig:sim_two_tweezer_splitter}
\end{figure*}

This splitting scheme starts with two tweezers, one holding the atoms and the other one empty, positioned at a large enough distance such that there is no tunneling between them. There are two parameters that vary in time: the potential difference between the tweezers (`detuning'), $\hbar \Delta$, and the tunneling rate, $J$. The former is controlled by adjusting the relative power between the beams, while the latter is determined by the distance $d(t)$ between the traps. At $t=0$, $J$ is set to zero and $\Delta$ is set to a small positive value, which means that the potential of the empty tweezer is shallower. The protocol is performed by moving the traps one towards the other, decreasing the distance $d(t)$ and increasing $J(t)$, and at the same time, the detuning parameter $\Delta(t)$ is lowered to zero. When the traps are closest to each other ($t_m$), $\Delta(t_m)$ is zero. Then, the distance $d(t)$ is increased again, while maintaining $\Delta(t)=0$.

The protocol described above splits the atomic wave function evenly between the two traps. To show this, we employ a tight binding model. For simplicity, we consider only one vibrational eigenstate in each tweezer, $\ket{\varphi_i}$, with energy $E_i$, where $i={1,2}$ identifies the tweezer. The Hamiltonian of this system, in the rotating wave approximation, can be written as
\begin{equation}
H=\hbar \Delta \ket{\varphi_2}\bra{\varphi_2}+\hbar \frac{J}{2}\ket{\varphi_2}\bra{\varphi_1}+h.c.\, \, .
\end{equation}
We can describe the state of the system using a Bloch vector ${\vec{v}=\left(\langle \sigma_x \rangle,\langle \sigma_y \rangle,\langle \sigma_z \rangle\right)}$, where $\sigma_i$ are the Pauli matrices operating in the two-dimensional subspace of $\{\varphi_1,\varphi_2\}$. The dynamics of the system is given by the optical Bloch equation: {$\dot{\vec{v}}=\mathcal{P}\times \vec{v}$}, where $\mathcal{P}=(J,0,\Delta)$ is the torque vector around which $\vec{v}$ performs precession. The initial state is $\ket{\psi}=\ket{\varphi_1}$. The initial detuning is chosen $\Delta\approx\omega_0$, where $\omega_0$ is the tweezer oscillation frequency. This choice is made to have the largest possible initial $\Delta$ before eigenstates with different vibrational numbers cross.

These initial conditions correspond to the Bloch and torque vectors being parallel, each pointing towards one of the poles. When the tweezers are gradually brought closer and the detuning is changed $\Delta\rightarrow 0$, the torque vector rotates to the equatorial plane, and the Bloch vector follows adiabatically. The scheme ends with a gradual decrease of $J\rightarrow 0$, leaving $\vec{v}$ in the equatorial plane. This means the wave function is $\ket{\psi}=\frac{1}{\sqrt{2}}\left[\ket{\varphi_1}+\ket{\varphi_2}\right]$, as desired. Importantly, because the process is adiabatic, it works with any initial vibrational eigenstate that fulfills the adiabatic condition. However, it is important to note that due to the non-harmonic nature of the potential, as more eigenstates are occupied, parameter adjustments become necessary to account for the smaller energy differences between higher eigenstates. In addition, the process should be executed at a slower pace to ensure adiabatic conditions are met. 

To evaluate the effectiveness of this splitting approach beyond the two-level approximation, we employed numerical solutions of the time-dependent Schrodinger equation using the split-step Fourier method \cite{Weideman1986}. Because the coupling is predominately in the radial direction, we model the system in one-dimension. Nonetheless, we confirmed our findings in a two-dimensional setting. 
 
\begin{figure}
	\centering
\includegraphics[width=1\linewidth]{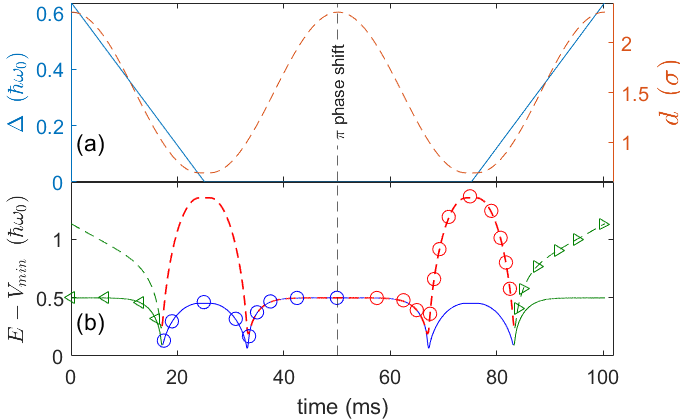}
\caption{\textbf{Two tweezers splitting and recombining process with a $\pi$ relative phase shift between the arms.} Simulation of traps with $\sigma$ = 1.3$\mu$m and depth of 115$\mu$K (a) The detuning between the tweezers $\Delta$ (solid blue line, left axis) and the separation between the tweezers centers $d$ (dashed red line, right axes) as a function of time. (b) The energies of the ground state (solid line) and the first excited state (dashed line) versus time. The color represents the wavefunction shape: green for an atom localized in a single trap (either left or right), blue (red) for a balanced (anti) symmetric splitting. The markers depict the evolution of the energy expectation value. Left (right) pointing triangles represent a wave packet localized on the left (right) tweezer, and circles represent a wave packet split between the tweezers. The dynamics depicted by the markers is of an atom initially in the left trap, then adiabatically following the ground state until it is split between both traps with a symmetric wavefunction. At the middle of the process, a relative $\pi$ phase shift is added between the interferometer arms, causing the wavefunction to become anti-symmetric. The atom proceeds to follow the excited state until it is localized in the right tweezer after the recombination. Similarly, if no phase shift is applied, the atom ends in the lower branch, which corresponds to the left tweezer.}	\label{fig:two_teezers_spectrum}
\end{figure}

The potential, which consists of two Gaussian tweezer beams, is given by:
\begin{equation}
V(x,t) = -V_0 \left[e^{-2\frac{x-d\left(t\right)/2}{\sigma^2}}+\Big(1-\Delta\left(t\right)\Big) e^{-2\frac{x+d\left(t\right)/2}{\sigma^2}}\right] \ \ ,
\end{equation}
where the time-dependent parameters are given by
\begin{align*}
d(t) =& \frac12 (d_{max}+d_{min}) + \frac12(d_{max}-d_{min})\cos \left(2\pi t/T\right)\\
\Delta(t)=&
\begin{cases}
        \Delta_{max}(1-2t/T) & \text{if }  t<T/2\\
        0 & \text{if } t \ge T/2
    \end{cases} \ \ ,
\end{align*}
where $T$ is the process duration (see Fig.~\ref{fig:two_teezers_spectrum}a). The initial wave-function is taken to be a specific eigenstate of one of the tweezers, calculated by numerical diagonalization of the Hamiltonian of that specific potential. There are two observables we calculate at the end of the simulated process; the probability to be in a given tweezer and the overlap fidelity of the wave function with the initial eigenstate. The overlap fidelity is given by $f_{right (left)} = \int_{-\infty}^{\infty} \psi^*(x)\varphi_{2 (1)}(x) \,dx$. The time steps and spatial resolution in the simulation are chosen to ensure the convergence of these observables. We optimize the process parameters to get as close as possible to unit fidelity and a probability of $1/2$ in each tweezer. 

The simulation results are shown in Fig.~\ref{fig:sim_two_tweezer_splitter} and Fig.~\ref{fig:two_teezers_spectrum}. Fig.~\ref{fig:sim_two_tweezer_splitter} demonstrates that the relative phase between the interferometer arms before recombining is indeed correlated with the population difference between the output tweezers. It also shows (second and last panel) that the interferometric loop works the same for both ground and excited vibrational states. Remarkably, we find that the splitting scheme works successfully even when the minimal distance between the tweezers is small enough such that there is no barrier and only a single minimum to the combined potential. In this regime, the tight-binding approximation does not hold, and the notion of tunneling needs to be reconsidered. The reason why the splitting scheme still works is because it is based on adiabatic following which can be generalized to merged potentials. Initially, when the tweezers are far and the occupied tweezer has a lower energy, the occupied state is essentially identical to the eigenstate of only a single tweezer. Then, when the tweezers are brought closer and the detuning is reduced to zero, this state evolves adiabatically to the symmetric state of the two tweezers. This adiabatic passage is protected by an avoided level crossing that opens a gap, which evolves from $2 \hbar J$, when the tweezers are only weakly coupled, to $\hbar \omega_0$, when they overlap ($\omega_0$ is the harmonic trapping frequency of the combined potential). Note that if the initial $\Delta$ is set to a negative instead of a positive value, the adiabatic following will end in the anti-symmetric wave function. In this case, there will be a $\pi$ phase between the interferometer arms.

Once the splitting process is over and the two output tweezers are taken apart, the gap between the symmetric and anti-symmetric states closes. This fact is crucial for the correct operation of the interferometer, since the phase shift between the atomic wave packets is translated into a specific mixing between the degenerate symmetric and anti-symmetric states of the two tweezer arms. Then, the time-reversed version of the splitting process achieves the coherent combining, where the differential phase shift becomes the relative population between the two tweezers exiting the combiner (see Fig.~\ref{fig:two_teezers_spectrum}b). Importantly, each of the symmetric and anti-symmetric states evolves adiabatically and eventually becomes a state localized in the left or right tweezer. This evolution maps the phase difference to population difference.

\subsection{Three-tweezers atomic splitter-combiner}\label{sec:Three-tweezers atomic splitter}
In this splitting scheme, we employ three tweezers as input and output ports. The atomic wave packet is confined initially to the central tweezer, located at $x=0$. Two empty tweezers are centered at a distance of $x=\pm d$, chosen such that the tunneling rate is negligible. Their energy detuning is set to $\Delta\approx -\omega_0$. The sequence proceeds by gradually bringing the two external tweezers closer to the central one while increasing $\Delta\rightarrow0$. At the minimal distance, the two external tweezers reverse their velocity and $\Delta=0$. Then, the external tweezers get farther away while $\Delta\rightarrow\omega_0$. After the process is done, the central tweezer is empty, while the atomic wave packet is in a balanced superposition of the two external tweezers. The time reversal of the process acts as a coherent combiner. 

To explain how this scheme works, we once again invoke the mapping to an effective two-level system ${\ket{\varphi_1},\ket{\varphi_S}}$, where $\ket{\varphi_1}$ is an eigenstate of the central tweezer and $\ket{\varphi_S}$ is the symmetric superposition of the two corresponding eigenstates in the external tweezers. The coupling $J(t)$ is again controlled by the distance between the tweezers, and $\hbar\Delta(t)$ is the energy detuning between the central and external tweezers. At the start of the process, the initial state is $\ket{\varphi_1}$, which means the initial Bloch vector points toward one of the poles. The torque vector starts by pointing towards the south pole, $\mathcal{P}=(0,0,-\omega_0)$,  then it gradually changes towards the equatorial plane, $\mathcal{P}=(J,0,0)$, crossing at the process midpoint, and continuing towards the north pole, $\mathcal{P}=(0,0,\omega_0)$. The Bloch vector follows adiabatically and ends pointing towards the opposite pole from where it started. This means the final state is $\ket{\varphi_S}$, which is the desired output state of the splitter.

\begin{figure}
	\centering
	\includegraphics[width=\linewidth]{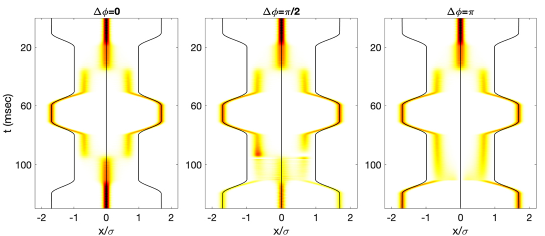}
	\caption{\textbf{Numerical simulation of the three-tweezer splitter-combiner.} The brightness of the color represents the probability distribution, $|\psi(x)|^2$ (darker is higher). The initial state is an atom in the ground vibrational state of the central tweezer. After splitting, the atom is in an equal superposition of the two external tweezers, and the central tweezer is empty. We have added various relative phase shifts between the arms in the middle of the process to demonstrate the operation of the combiner. This simulation assumed $^{40}K$ atom, and tweezers with a Gaussian waist of $\sigma=1.3\mu$m and a depth of $23\mu$K.}
	\label{fig:sim_three_tweezers_splitter}
\end{figure}

In the recombination, the observable is the population difference between the central and external tweezers. To see this, consider a state before combining, $\ket{\psi}=\ket{R}+e^{i\phi}\ket{L}$, where ${\ket{R},\ket{L}}$ are localized eigenstates in the right and left tweezers and $\phi$ is the relative phase shift. We can rewrite this state as $\ket{\psi}=e^{i\frac{\phi}{2}}\left[\cos(\frac{\phi}{2})\ket{\varphi_S}+\sin(\frac{\phi}{2})\ket{\varphi_{AS}}\right]$, where $\ket{\varphi_{S/AS}}$ are the symmetric and anti-symmetric superpositions of ${\ket{R},\ket{L}}$. Due to symmetry, the combiner couples only the symmetric states, $\ket{\varphi_S}$ and the eigenstate in the central tweezer ($\ket{\varphi_1}$). The recombination process follows the adiabatic following in reverse, transforming the state to $\ket{\psi}=e^{i\frac{\phi}{2}}\left[\cos(\frac{\phi}{2})\ket{\varphi_1}+\sin(\frac{\phi}{2})\ket{\varphi_{AS}}\right]$. The probability of finding an atom in the external tweezers is $\sin^2(\frac{\phi}{2})$, from which we can determine the phase. In Fig. \ref{fig:sim_three_tweezers_splitter} we plot the results of 1D numerical simulations with typical experimental parameters. The simulations demonstrate that a relative phase shift between the two external interferometer arms is indeed mapped to relative population difference between the central tweezer the two external tweezers.

\begin{figure}[!ht]
	\centering
\includegraphics[width=\linewidth]{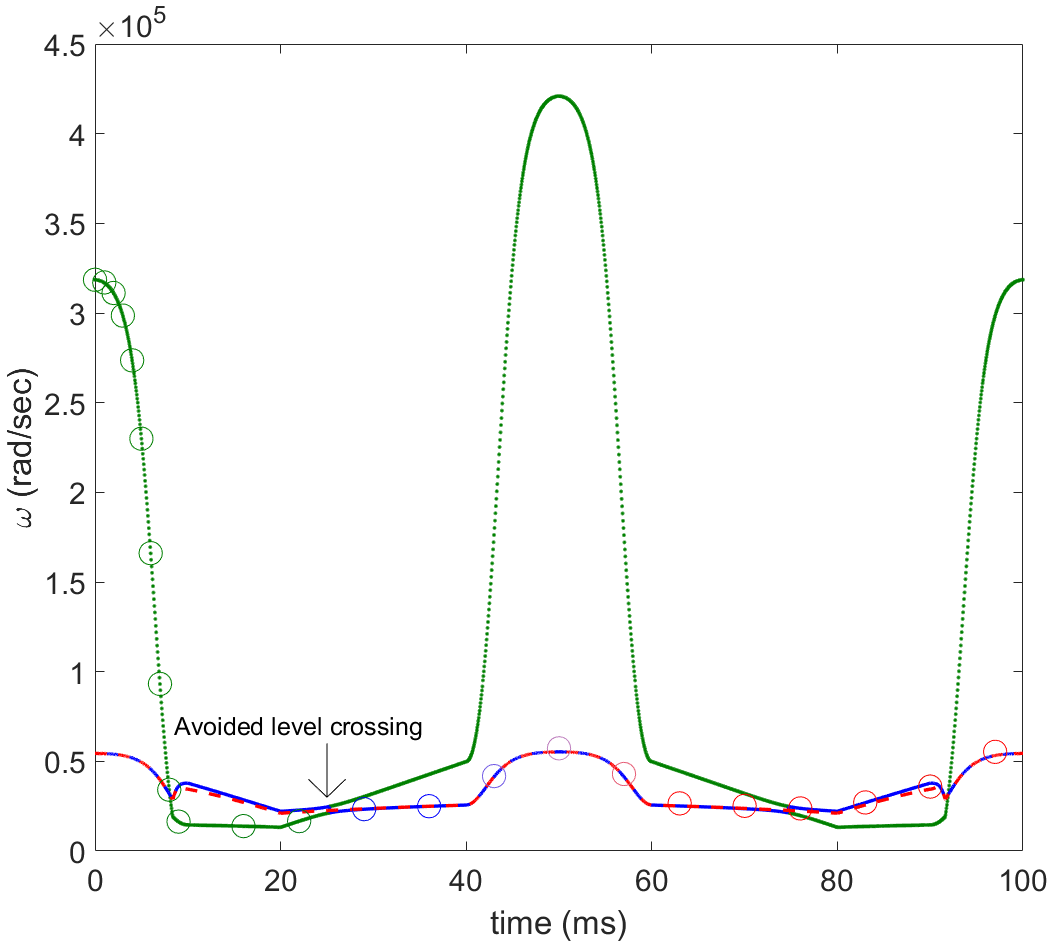}
	\caption{\textbf{Adiabatic following in the three tweezers splitter-combiner.} The lines show the three lowest eigenenergies (divided by $\hbar$) throughout the process, while the circles follow the evolution of the atomic energy expectation value. The color represents whether the wave packet is localized on the central tweezer (green, dotted), or if it is in a symmetric (blue, solid) or anti-symmetric (red, dashed) superposition of the side tweezers. A notable feature of the spectrum is the avoided level crossing, occurring at approximately 25 ms, which allows the atom to move adiabatically from the central tweezer to the side tweezers. The splitter output state is symmetric since the potential is symmetric. After time $50\text{[ms]}$, the process is time reversed. If the atom remains in a symmetric state, its wave-packet follows back into the central trap. In this simulation, however, we apply an additional differential phase shift of $\pi$ between the two arms in the middle of the process, which is indicated by the transition of the $\circ$ markers' color from blue to red. In this case, symmetry dictates adiabatic following to the anti symmetric superposition of the two side tweezers.}
	\label{fig:states_of_3_trap_interferometer}
\end{figure}

Similar to the two-tweezer splitter, the three-tweezers scheme also functions effectively even when the traps are partially merged. This phenomenon can be explained using the adiabatic theorem. In Fig.~\ref{fig:states_of_3_trap_interferometer}, we depict the evolution of the relevant eigenenergies during the process of splitting and recombining. Due to the avoided crossing effect \cite{demkov2007neumann}, the eigenstates $\ket{\varphi_1}$ and $\ket{\varphi_S}$ are connected by a single smooth branch, along which the state evolves adiabatically during the splitting stage. In the combining stage, the symmetric state of the two external tweezers transitions from $\ket{\varphi_S}$ back to $\ket{\varphi_1}$, while the anti-symmetric state $\ket{\varphi_{AS}}$ remains unchanged. It is crucial that the detuning parameter is kept $\Delta<\hbar\omega_0$ to prevent coupling between the anti-symmetric eigenstate and the second excited level of the central tweezer, which is also anti-symmetric. This requirement is essential to ensure that when a relative $\pi$ phase difference exists between the interferometer arms, the atom remains in the external tweezers.

There are two main advantages to the three-tweezer splitter-combiner. First, the process does not need to be fine-tuned to end at $\Delta=0$. It is therefore simpler to implement and more robust. Second, the process has an error indicator; After splitting, the central tweezer should be empty. Thus, a measurement of the population at the central tweezer indicates the fidelity of the splitting process without perturbing the interferometer arms. In the recombination process, we note that the population in the two external tweezers should be equal since they are in the anti-symmetric state. Therefore, the error indicator, in this case, is any population difference between the external tweezers. The two error indicators allow to reject experimental runs that were severely affected by noise, thus increasing the interferometer precision.

\section{Sensitivity and precision}\label{sec:Sensitivity_estimation}

The evolution of the wave function is given by $\ket{\psi(t)}=e^{-i\nicefrac{S_\Gamma}{\hbar}}\ket{\psi(t=0)}$, where $S_\Gamma$ is the action, defined by the integration over the Lagrangian along the classical path $\Gamma$: $S_\Gamma=\int_\Gamma \mathcal{L}[\mathbf{r},\dot{\mathbf{r}}]dt$ \cite{Storey1994,Cronin2009}. In our interferometer, defined by two paths $\Gamma_1$ and $\Gamma_2$ for the two arms, the relative phase shift acquired by an atom is therefore {$\Delta \phi=\frac{1}{\hbar} \int_{\Gamma_2} \mathcal{L}[\mathbf{r},\dot{\mathbf{r}}]dt-\frac{1}{\hbar} \int_{\Gamma_1} \mathcal{L}[\mathbf{r},\dot{\mathbf{r}}]dt$}. To assess the sensitivity of the tweezer interferometer, we consider the simplest scenario where the atom is split symmetrically and separated to a distance $h$, where it is held at rest, and then recombined. We assume the movements are symmetric and short compared to the total measurement duration, $T$, and therefore do not include their contribution to the phase difference. 

Let us consider for simplicity that the interferometer is subjected to a uniform acceleration $a$ (e.g., gravity) aligned parallel to the line connecting the two tweezers. Then, we can write $\mathcal{L}=-m\cdot x\cdot a$, where $m$ and $x$ are the atom's mass and position along this line. The phase difference is $\Delta \phi=\frac{m\cdot h\cdot a}{\hbar}T$. $T$ is ultimately limited by the lifetime of atoms in the tweezers, which can be many tens of seconds. Using acousto-optics deflector (AOD) technology, the distance between tweezers can be tuned up to hundreds of micrometers. The distance can be further increased if the tweezers are generated by two separate AODs, steered by two piezo-controlled mirrors, and then combined with a beam splitter. This optical scheme allows for precise control at short distances using the AOD and reaching large distances with the piezo mirrors. Thus, the distance between tweezers is only limited by the objective field of view. We estimate that it should be possible to reach a separation of 10-50 mm. Taking $T=10$ s and $h=10$ mm, we obtain that $^{40}$K atoms will acquire a phase shift of $\Delta \phi\approx 6.2 \cdot 10^8$ rad due to earth's gravitational acceleration ($a=g$). By increasing the distance between the tweezers to 50 mm and the waiting time to a minute, the phase difference is increased to $\sim 2 \cdot 10^{10}$ rad. 

Let us compare these numbers to the conventional Kasevich-Chu atomic interferometer. There, the phase shift is given by $\Delta \phi=k_\text{eff} g T^2$, with $k_\text{eff}$ being the effective wave-vector of the momentum kick given to split between the two arms \cite{Dimopoulos2007}. Taking as typical numbers $k_\text{eff}=\frac{4\pi}{780nm}$ and $T=1$ s \cite{Rosi2014}, one obtains $\Delta \phi\approx 1.6 \cdot 10^8$ rad for the gravitational accelerations. More advanced versions of the Kasevich-Chu interferometer can impart a larger momentum kick of few tens to a hundred $\hbar k$ (k in the Raman laser wave vector), but this comes with a price of large sensitivity to wave front distortions and phase noise of the Raman beams  \cite{PhysRevLett.107.130403}. 

Ideal operation conditions of the interferometer require that the two tweezers have the same intensity. Thus, as long as intensity fluctuations are common to both tweezers, they do not introduce relative phase noise. Relative fluctuations, on the other hand, lead to a relative phase shift, thereby impairing the interferometer's operation. We now consider a fundamental source of such relative intensity fluctuations which is the shot noise of each beam. Let us denote the peak power of each tweezer as $P_0$. The average number of photons in each tweezer during the experiment is $N=\nicefrac{T P_0}{\hbar\omega_t}$, where $\omega_t$ is the tweezer laser angular frequency. The relative phase noise is given by $\nicefrac{\delta \phi}{\Phi}=\sqrt{2}\cdot\nicefrac{\delta N}{N}$, which for a shot noise is $\sqrt{2} (N)^{-1/2}$. Importantly, here $\Phi$ is the phase acquired in each tweezer due to the optical potential and not due to the external potential. Thus, we obtain 
\begin{equation}
    \Big(\frac{\delta \phi}{\Phi}\Big)_{\text{shot noise}}=\sqrt{\frac{2\hbar\omega_t}{T P_0}} \ \ .
\end{equation}
As an example, we consider a $^{40}$K atom held for 10 s in a tweezer interferometer with a wavelength of $\lambda=2 \mu m$, a Gaussian waist of 1.7 $\mu m$, and a power of \mbox{120 $\mu$W} in each arm. This yields a tweezer depth of approximately 2 $\mu$K, for which $\nicefrac{\delta \phi}{\Phi}\approx 1.3\cdot10^{-8}$ and $\delta \phi \approx 34$ mrad. For this trap, the ground state width in the radial direction is approximately $0.256\,\mu$m, which defines the ultimate spatial resolution of the interferometer. It is important to note that the number of bound states in this tweezer potential is much larger than 100, as we have verified numerically. Hence, it can hold more than 100 fermionic atoms in different vibrational states. Clearly, mitigating all the technical noise sources such that the limiting factor is the shot noise is an experimental challenge. 

Next, we examine the sensitivity of the splitting and recombining stages to noise. We assume that the differential noise between the tweezers has reached the shot noise limit. By solving numerically the time-dependent Schrodinger equation, we found that shot noise has no measurable effect on the performance of the splitter or combiner. However, common mode fluctuations of the tweezer intensity may still exist. To asses their impact, we run simulations where we introduce random variations with zero mean, $V_n(t)$, to the tweezer amplitudes. $V_n(t)$ is characterized by white noise with spectral density $S(f)=\eta 2\hbar\omega_t P_0$, which is $\eta$ times larger than the shot noise. In most cases, the amplitude of the common-mode noise will surpass the shot noise significantly; A typical number we use is that it will be stronger by around 70 dB. Additionally, we assume that this noise is uniformly applied to the tweezers.

For every noise realization, we perform a complete simulation of the interferometer loop, and determine the probabilities of locating the atom in each of the output arms. For a three tweezer splitter, we also evaluate after the splitting stage the probability to find an atom in the central trap, and accordingly randomly determine if the atom is there. Such an event signals that the splitting stage has failed, in which case the numerical experiment is declared unsuccessful. If the atom is not drawn to be at the center, we project the wavefunction to the outer tweezer arms, maintaining their relative probabilities and phase, and proceed to simulate the rest of the interferometer loop. The relative phase is determined by adding a deterministic phase which is scanned. Therefore, we repeat the complete simulation 4000 times, each time with a different noise realization and different deterministic relative phase in the range $0-2\pi$ between the arms. The results of these simulations were stored and used as a lookup table to calculate the probability outcomes after the combiner, as described below.

\begin{DIFnomarkup}
\begin{table}[!ht]
	\begin{tabular}{|>{\centering\arraybackslash}p{1.2cm}|>{\centering\arraybackslash}p{1.5cm}|>{\centering\arraybackslash}p{1.5cm}|>{\centering\arraybackslash}p{1cm}|>{\centering\arraybackslash}p{2.5cm}|>{\centering\arraybackslash}p{2.5cm}|>{\centering\arraybackslash}p{2.2cm}|}
		\hline
		scenario number & no. of repetitions & no. of atoms in each run & T [sec] & total run-time \\
		\hline \hline
		1 	& 10 & 1  & 10 	& $\sim$50min \\
		\hline
		2 	& 10 & 10  & 10 & $\sim$50min \\
		\hline
		3   & 285 & 100 & 10  & $\sim$24hr\\
		\hline
		4   & 96 & 10 & 40  & $\sim$24hr\\
		\hline
		5   & 96 & 100 & 40  & $\sim$24hr\\
		\hline
	\end{tabular}
	\caption{\textbf{Different types of experimental scenarios with a tweezer-based interferometer.} The scenarios vary in the number of repetitions per deterministic phase, number of atoms in the tweezer per run, the duration of each run (T), and the total duration of the experiment.} \label{tab:experiments}
\end{table}
\end{DIFnomarkup}

We consider five different types of interferometry experiments that vary in their probing duration (T), the number of atoms per run, and the number of repetitions per deterministic phase (see Table \ref{tab:experiments}). Two of the scenarios are 'short', namely can be completed in less than an hour, and three are 'long', taking around a day of data integration. In each scenario, we choose to have 20 equally spaced deterministic phases to simulate a `fringe scan' and determine the phase shift between the arms, which is due to the physical phenomenon under investigation. For each of these 20 deterministic phases, a random noise of 34 (68) mrad is added to account for shot noise for T=10 sec (T = 40 sec). This is justified since after the splitting and  during the measurement time, the tweezers' amplitudes will be lowered to 2 $\mu K$ and then increased back for the combiner stage. If multiple atoms are involved in a run, they all experience the same noise realization. In each run, we select the closest phase out of the 4000 realizations and use its calculated quantum wave function to determine the output probabilities. Finally, we use these probabilities to randomly assign an exit port for each atom. We then fit the numerical data, extract the phase in the presence of noise and determine the error relative to the known physical phase. This procedure enables us to obtain the expected accuracy of the interferometer in various realistic scenarios.

\begin{DIFnomarkup}
\begin{table*}[!ht]
	\begin{tabular}{|>{\centering\arraybackslash}p{1cm}|>{\centering\arraybackslash}p{1.1cm}|>{\centering\arraybackslash}p{1.3cm}|>{\centering\arraybackslash}p{2.5cm}|>{\centering\arraybackslash}p{2.5cm}|>{\centering\arraybackslash}p{2.5cm}|>{\centering\arraybackslash}p{2.5cm}|>{\centering\arraybackslash}p{2.5cm}|}
		\hline
		$\eta$ & splitter & fail ratio& \multicolumn{5}{c|}{phase uncertainty [mrad]}\\
		\cline{4-8}
		&type&[\%]&scenario 1&scenario 2&scenario 3&scenario 4&scenario 5\\
		\hline \hline
		1 & II & --- & 87  & 28 &  2.4  & 9 & 3.8\\
		\hline
		$10^7$ & II & --- & 90  & 30 &  2.3 & 10.1 & 3.8 \\
		\hline
		$5\cdot 10^7$ & II & --- & 193  & 96 &  36 & 44  & 37\\ 
		\hline
		$10^8$ & II & --- & 230  & 180 &  173 & 169 & 162\\  
		\hline
		1 & III & 0.2 & 85  & 28 &  2 & 9 & 3.5\\
		\hline
		$10^6$ & III & 1.6 & 97  & 30 &  3.2 & 9.5  & 4.4 \\
		\hline
		$10^7$ & III & 13 & 164  & 89 &  60 & 61 & 60\\
		\hline
		$5\cdot 10^7$ & III & 32 & undefined  & undefined &  undefined &  undefined& undefined	 \\  
		\hline
	\end{tabular}
	\caption{\textbf{Error estimation for the tweezer atomic interferometer -- simulation results.} The table shows the results with several experimental scenarios, as detailed in Table \ref{tab:experiments}. We simulate both the two tweezers (denoted as type II) or three tweezers (type III) splitter-combiner schemes. The fail ratio, available only for type III splitter-combiner, is the probability to find the atom in the central tweezer after the splitting stage.} \label{tab:noise2}
\end{table*}
\end{DIFnomarkup}

There are two main types of events that may occur during the measurement and lead to the collapse of the wave function. The first type involves a collision with an energetic atom or molecule from the remaining ambient gas within the vacuum chamber. This process is characterized by the so-called vacuum lifetime, typically ranging from tens to few hundreds of seconds in conventional ultra-high vacuum systems and in cryogenic vacuum systems it can even extend to thousands of seconds \cite{Schymik2021}. The second process involves spontaneous photon scattering from the trap light. The rate of this process for a single atom scales as $I/\Delta$, where $I$ represents the intensity of the laser, and $\Delta$ denotes its detuning from the strong atomic lines \cite{Grimm2000}. Therefore, it is advantageous to work with the largest possible detuning and the weakest trap laser intensity.

For our calculations, we opted for a wavelength of $\lambda=2 \mu m$ due to its compatibility with standard optical components and its availability with commercial lasers. For a tweezer with a Gaussian waist of 1.7 $\mu m$ and a depth of $2, \mu$K, as discussed earlier, the average time before a spontaneous photon scattering event for a $^{40}$K atom is approximately 217 seconds. Consequently, for a probing duration of $T=10$ seconds, the percentage of experimental runs in which scattering occurs in one of the interferometer arms is around $9\%$. This percentage increases to $31\%$ when $T=40$ seconds. It's important to note that in runs where a collapse occurs, it introduces only a bias signal without inducing fringe oscillations into the measurement. This is because the combiner stage results in an even split when initiated with a wave packet in only one of its incoming ports. Therefore, these events only moderately reduce the contrast but do not introduce systematic errors. Additionally, the rate of spontaneous scattering may be suppressed due to Pauli blocking by other fermionic atoms in the tweezer, a phenomenon observed in a large optical trap \cite{Busch1998, Margalit2021, Sanner2021, Deb2021}.

The results of the error estimation for the different scenarios are presented in Table \ref{tab:noise2}. The simulation shows that the two tweezer splitter-combiner loop is resilient to noise up to $\eta\approx 10^8$, while the three tweezer splitter-combiner is slightly more sensitive, becoming non-operational at around $\eta \approx 5 \cdot 10^7$. We find that errors due to uneven division of the waveform by the splitter-combiner are largely unaffected in a wide range of noise levels, as long as $\eta < 10^7$. Typical commercial lasers, however, have much lower $\eta$ values, ranging from $10^3$ to $10^4$. By comparing the results of scenario 1 and 2, the relative advantage of working with several atoms in a run is clearly demonstrated. By comparing the results of scenarios 3 and 5, we see that the uncertainty is similar, particularly at a large noise level, while the phase accumulation in scenario 5 is four times larger due to the longer probing duration. Therefore, we conclude that under this white noise model, and as long as the phase noise due to shot noise is small enough, the tweezer interferometer benefits from prolonging its probing duration.

The relative accuracy of the interferometer depends on the total relative phase induced by the investigated physical phenomenon. We return to the example of measuring the earth gravitational acceleration with $^{40}$K atoms (h=10 mm, T=10 sec), where $\Delta \phi\approx 6.2 \cdot 10^8$ rad. The phase uncertainty levels presented in Table \ref{tab:noise2} for scenarios 1-3 allow us to estimate that, as long as the relative intensity noise is shot noise limited, the relative accuracy could reach the $10^{-10}$ level in a wide range of common-mode noise levels by separating the wave packet for 10 seconds. By extending the probing time to 40 seconds, this accuracy can be further improved below the $10^{-11}$ level. These results and the low sensitivity to common-mode noise are very promising for the actual implementation of the tweezer interferometer.

\section{Applications}\label{sec:Applications}
The proposed tweezer-based atomic interferometer has the potential to revolutionize many fields where an extremely sensitive force detection with sub micrometer resolution is needed. In this section we discuss two concrete examples -- measurement of the gravitational constant and measurement of surface forces due to quantum vacuum fluctuations (Casimir-Polder force).

\subsection{Measurement of the Casimir-Polder force}
One of the most striking predictions of quantum field theory is the existence of forces between two objects in empty space due to the vacuum fluctuations of the electromagnetic field between them \cite{Dalvit2011}. These forces were first studied by Hendrick Casimir in 1948 for two surfaces \cite{Casimir1948} and later by Casimir and Polder for an atom and a surface \cite{Casimir1948a}. The full calculation of these forces using quantum electrodynamics (QED) is quite complex, but analytical results can be obtained for certain limiting cases \cite{Scheel10.1088/1367-2630/8/10/2372008}. When the atom is very close to the surface, the interaction can be described as an attraction between the fluctuating atomic dipole and its mirror image. The Casimir-Polder (CP) potential in this case, which is also referred to as Lennard-Jones or van der Waals potential, scales as $U_{CP}\propto 1/z^3$, where $z$ is the distance to the surface. In the opposite limit, called the retarded limit, the potential scales as $U_{CP}\propto 1/z^4$. The transition between these two regimes occurs at a typical lengthscale of  $l\sim100$ nm. The exact CP force depends on the surface electrical properties, roughness, and temperature. Their precise measurement is important to test approximation methods in QED and as a means to understand material properties. Casimir forces are generally small, but they have a significant impact at the nanoscale, making them crucial for nano-technology applications, specifically micro-electro-mechanical systems (MEMS). Additionally, a thorough understanding of these forces is necessary before searching for new physics beyond the standard model at very short length scales  \cite{Harber2005,Onofrio2006,Klimchitskaya2023}.

The force between two surfaces has been measured with increasing precision since the late 1990s, with good agreement between experiments and theory \cite{Lamoreaux1997,Mohideen1998,Chan2001,Chan2001a,Decca2003}. The CP force  between an atom and a surface has been measured with increasing accuracy since 1975 using various techniques, including atomic-beam-deflection experiments \cite{Shih1975,Sukenik1993}, laser spectroscopy of atoms near a wall  \cite{Sandoghdar1992}, interaction with a diffraction grating \cite{Perreault2005}, quantum reflection experiments \cite{Shimizu2001,Druzhinina2003,Pasquini2004}, ultracold atoms bouncing off of an atomic mirror \cite{Landragin1996,Bender2010,Fuchs2018}, and measurement changes of the oscillation frequency of trapped $^{87}$Rb Bose-Einstein condensates near a surface \cite{Harber2005,Obrecht2007}. There are significant discrepancies between theory and experimental results, mainly due the difficulty in measuring directly the force acting on a single atom. Most of the experiments relied on measuring the CP force using kinematic effects or spectroscopic probes. A notable exception is the experiment by Perreault and Cronin, where a diffraction-based atomic interferometry was used and the Casimir-Polder potential was manifested as an additional phase shift of the interference fringes due to the interaction with the grating at close proximity  \cite{Perreault2005}. However, the signal was very weak and had large uncertainty, mainly due to the short interaction time.

\begin{figure}[!ht]
	\centering
	\includegraphics[width=1\linewidth] {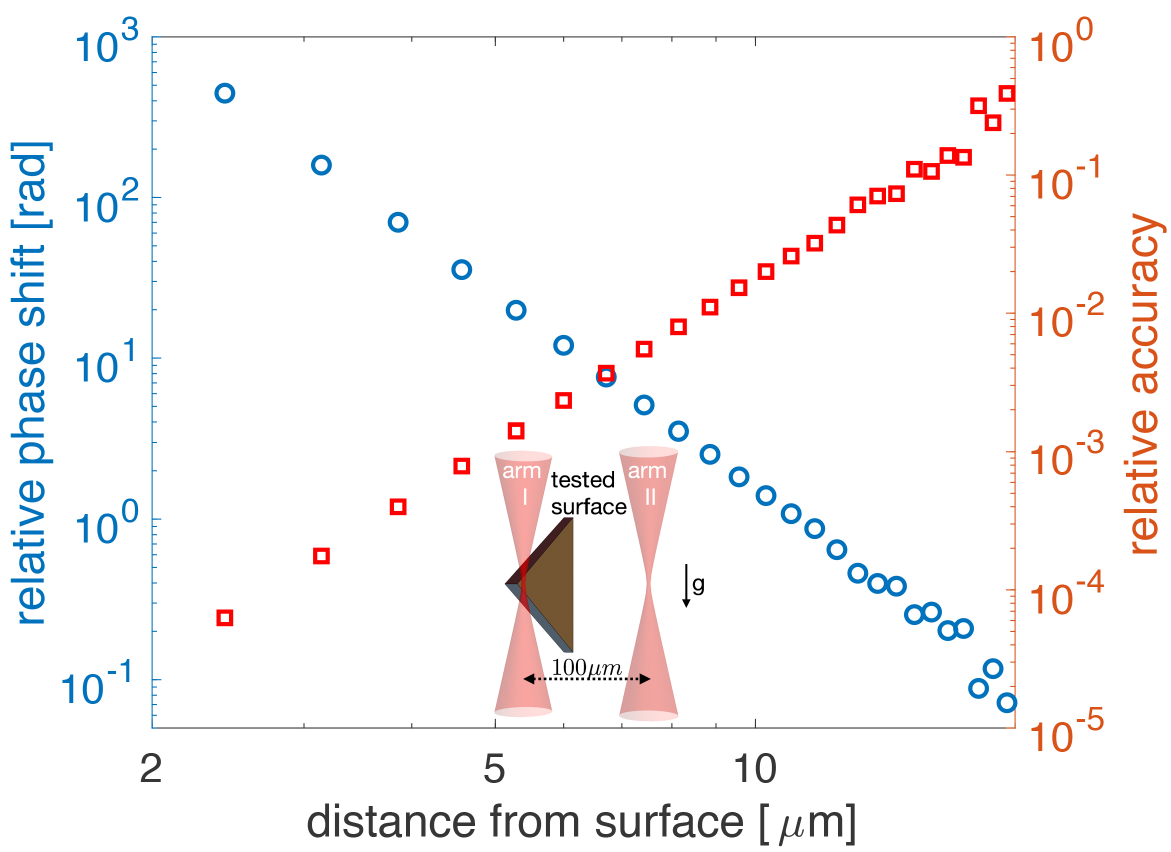}
\caption{\textbf{Measurement of the Casimir-Polder force using a tweezer atomic interferometer.} The experiment we consider is depicted in the inset: one arm being at a short distance from a metallic surface (x-axis) and a second reference arm is positioned $100\mu$m away. The calculated relative phase shift between the arms due to the retarded CP potential is shown as blue circles (left y-axis), and the corresponding relative accuracy using scenario 2 in Table \ref{tab:experiments} is shown as red squares (right y-axis). The number of points in this graph was chosen such that it would be possible to acquire the data in approximately 24 hours.}
\label{fig:CP}
\end{figure}

The tweezer-based atomic interferometer is perfectly suited to measure the phase shift induced by the CP interaction thanks to its ability to precisely position the atomic wave packets for a long duration. The idea of the measurement is depicted in the inset of Fig. \ref{fig:CP}. One tweezer arm will be positioned close to a surface, where it will acquire a phase shift due to the Casimir-Polder potential, while the second reference arm will be 100 micrometers away, where the potential is negligible. In Fig. \ref{fig:CP}, we present a calculation of the relative phase shift acquired due to the CP potential versus the distance of the first tweezer from the surface. The phase shift is calculated in the retarded limit $U_{CP}(z)=-C_4/z^4$, since the tweezer position $z\gg l$, where $l \approx 118$ nm \cite{Friedrich2002}. The constant $C_4=1.64 \cdot 10^{-55}$ J m$^{4}$ was taken from Ref. \cite{Friedrich2002}, where it was calculated for $^{40}$K atoms near a metallic surface. The effect has a similar order of magnitude with a dielectric surface \cite{DZYALOSHINSKII1992}. We have also verified that the non-additive effect of the tweezer on CP potential for our tweezer parameters is negligible \cite{Fuchs2018}. In Fig. \ref{fig:CP} we also present the expected relative accuracy of this measurement, assuming for each point the operating conditions of scenario 2 in Table \ref{tab:experiments}. Our calculation shows that with the tweezer AIF, one can map the CP potential in the region $1\ll z<20\mu$m with very high precision. For example, at a distance of $5\mu$m, we estimate a measurement with relative precision of $\sim 10^{-3}$. Most importantly, the suggested interferometric measurement can yield a direct, precise, and model-independent determination of the CP force.

One practical issue to consider is to have a design that ensures that the sample does not partially block the tweezer's Gaussian beam. Our solution, plotted in the inset of Fig. \ref{fig:CP}, is to shape the sample as a triangle. It is based on the assumption that at a given atom-surface distance $z$, a surface area of approximately $10z \times 10z$ is large enough to approximate sufficiently well the infinite surface limit. The triangular shape of the sample allows to move the tweezer towards the base of the triangle as it is taken farther from the surface. This way, the ratio between the distance and the relevant surface area is maintained, as well as a clear solid angle. 

\subsection{Measurement of the gravitational constant}
The value of Newton's gravitational constant, $G$, determines the strength of the gravitational force between two masses, and its precise value is crucial for a wide range of applications, including the study of celestial bodies and the prediction of the orbits of satellites and planets. There have been many attempts to measure $G$ over the years using a variety of techniques, including torsion balances, Cavendish balances, and spacecraft tracking. Still, $G$ is the least known of all fundamental constants, with a slow improvement in its accuracy. The value of G provided by the 2018 Committee on Data of the International Science Council (CODATA) has a relative uncertainty of $2.2 \cdot 10^{-5}$ \cite{Tiesinga2021}, considerably larger compared to other constants, such as the fine structure constant ($1.5\cdot 10^{-10}$), the electron mass ($3\cdot 10^{-10}$), or the vacuum electric permittivity ($1.5\cdot 10^{-10}$). The reason for the relatively high uncertainty in G is the weakness of gravity compared to the other forces. Most worrying is the inconsistencies between measurements done using different techniques, or even among those done with the same method.

In 2007, the M. Kasevich group published the first measurement of Newton's gravitational constant using atom interferometry \cite{Fixler2007}. Their experiment used a dual interferometer setup to eliminate the influence of Earth's gravity while being sensitive to the gravitational force of a nearby 540kg source mass. This groundbreaking experiment had a relative accuracy of $4\cdot 10^{-3}$ and the value it reported for $G$ was $1.0028$ higher than the recommended 2018 CODATA value. A second determination of $G$ with atom interferometry was reported in 2014 by G. Tino's group\cite{Rosi2014}. This experiment also employed a dual interferometer setup with a 516kg source mass. It had a smaller relative uncertainty of $1.5\cdot 10^{-4}$ and is the only atom interferometry measurement included in the CODATA 2018 determination of $G$. However, its reported value of $G$ is 2.4 standard deviations lower than the CODATA recommended value.

\begin{figure}[!ht]
	\centering
	\includegraphics[width=0.6\linewidth] {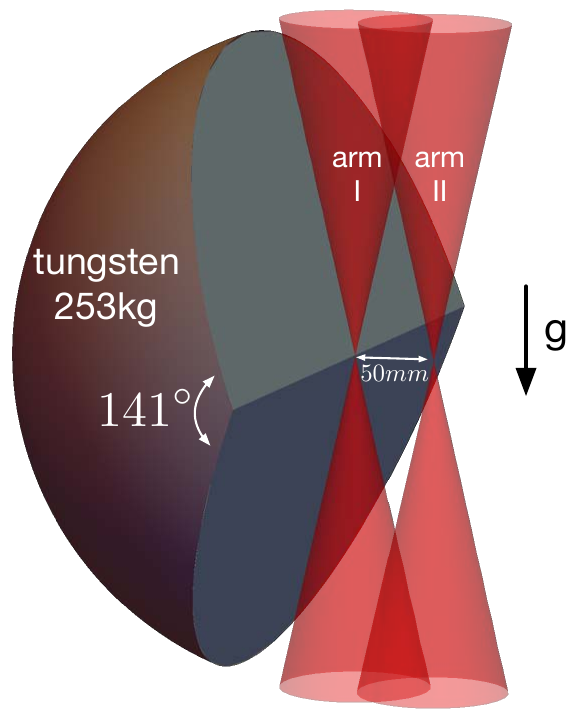}
	\caption{\textbf{Schematic (not to scale) of the proposed experiment to measure the gravitational constant.} The tweezer cone angle is determined by the waist of the Gaussian beam, which in this case is $\sigma=1.5\mu$m.}
	\label{fig:Big_G_exp}
\end{figure}

We suggest a different approach to measuring $G$ using tweezer atom interferometry. Our approach, shown in Fig. \ref{fig:Big_G_exp}, benefits from the ability to position the atoms near a test mass for a long duration, in a geometry that eliminates the effect of earth's gravitational field. We propose using a $\sim 253$kg tungsten source mass, which is shaped as sphere with a radius of 20cm cut at an angle of $141^\circ$. This shape was designed to avoid clipping the tweezers' Gaussian beams. The atomic wave packets in the two arms of the interferometer are positioned in a plane perpendicular to earth gravity that also contains the center of the cut sphere. This configuration cancels out the effect of Earth's gravitational potential in the interferometric measurement without a need for a dual interferometer setup. The wave packets will be placed at distances of 1 mm and 50 mm away from the center of the cut sphere, where the latter should be as large as possible. The experiment can be conducted with and without the source mass to eliminate systematic deviations. 

The interferometer sequence begins by splitting the atomic wave packet and moving the two tweezers to the two positions. The wave packets are held in these positions for a duration $T$ before moving back and recombined. The fringe scan can be done by changing the duration $T+\Delta T$, with $|\Delta T| < 20$ms. For $T=10$ sec, the phases accumulated by $^{40}$K atoms due to the sphere's gravitational potential are $\sim 798$ rad and $\sim 615$ rad for the short and long distances, respectively. With the estimated phase uncertainty of scenario no. 3 (see Tables \ref{tab:experiments} and \ref{tab:noise2}), and assuming that all the experimental parameters are known with high enough precision, G could be determined with a relative accuracy of $\sim 1.3 \cdot 10^{-5}$. A longer probing duration of $T=40$ sec (scenario 5) can reduce the uncertainty to $\sim 5 \cdot 10^{-6}$. Lastly, employing a heavier fermions, such as $^{171}$Yb, can reduce the uncertainty by the mass ratio, namely by a factor of $4.275$. The proposed experiment can determine the gravitational constant using a completely new approach with an uncertainty that has the potential to go below the current CODATA value.

\section{Summary}\label{sec:Summary}
We have presented a new approach to atomic interferometry which is based on ultracold fermions in reconfigurable optical tweezers. The interferometer key elements are the adiabatic splitter and combiner schemes, together with the ability to prepare many atoms in different vibrational states and use them together in a single run of the interferometer. Thank for the small size of the optical traps, the wave packets are positioned with sub micron precision and their path can be freely controlled. In particular, the wave packet can be held completely stationary for long probing duration. The remarkable advancements in optical tweezer technology, driven by its potential in quantum computation, have culminated in substantial progress in recent years, rendering it sufficiently mature for the practical realization of these concepts.

The unique capabilities of the tweezer interferometer can be transformative in many fields of science. We have discussed in detail the case of measuring the Casimir-Polder surface forces and determination of the gravitational constant. Another interesting application is to map the gravitational forces at short distances to test non-Newtonian gravity theories \cite{Newman2009,Yang2012,Ke2021}. The new interferometer can also be utilized to search for quantum gravity effects, e.g., detection of entanglement between two different configurations of an atom and a mechanical oscillator \cite{Carney2021}. Another promising application is to study material properties in condensed matter. Specifically, the tweezer interferometer can be used to measure forces of localized topological defects such as vortices or skyrmions. Furthermore, the interferometer's unprecedented sensitivity enables precise mapping of magnetic fields in proximity to surfaces.

\textit{Note added in proof.} Recently, we became aware of a related work \cite{Premawardhana2023}.

\begin{acknowledgments}
	We thank Yanay Florshaim, Elad Zohar, and Amir Stern for helpful discussions. This research was supported by the Israel Science Foundation (ISF), grant No. 3491/21, and by the Helen Diller Quantum Center at the Technion.
\end{acknowledgments}

\FloatBarrier

%

\end{document}